\documentclass[conference,a4paper]{IEEEtran}
\usepackage{mathtools, cuted}
\usepackage{lipsum, color}

\usepackage{footnote}
\usepackage[table]{xcolor}
\usepackage{times}
\usepackage{epsfig}
\usepackage{amsmath}
\usepackage{amsthm}
\usepackage{amsfonts}
\usepackage{graphicx}
\usepackage{amssymb}
\usepackage{amstext}
\usepackage{latexsym}
\usepackage{color,colortbl}
\usepackage{ifthen}
\usepackage{multirow}
\usepackage{verbatim}
\usepackage{array,tabularx}
\usepackage{arydshln}
\usepackage[mathscr]{euscript}
\usepackage{accents}
 \usepackage{cite}
\usepackage{hhline}
\usepackage{caption}
\usepackage{subcaption}
\usepackage{enumerate}
\usepackage{xcolor}
\usepackage{mathtools}
\usepackage{url}
\usepackage{xparse}
\usepackage{makecell}
\usepackage{varwidth}
\usepackage{bm}
\usepackage[bookmarks,colorlinks]{hyperref}
\usepackage{arydshln}
\usepackage{footmisc} 

\usepackage{comment}
\usepackage{wrapfig}

\usepackage{caption}
\usepackage{float}
\usepackage{booktabs}

\usepackage{mathtools}

\usepackage{booktabs}

\usepackage{float}

\usepackage{multirow}

\usepackage{acronym}

\newcolumntype{C}[1]{>{\centering\let\newline\\\arraybackslash\hspace{0pt}}m{#1}}

\usepackage[normalem]{ulem}

\usepackage{amsmath,pgfplots,amssymb}
\usepackage{subcaption,graphicx}
\usepackage{amsmath,dsfont}

\usepackage{algorithm}
\usepackage{algpseudocode}

\theoremstyle{definition}

\theoremstyle{definition}

\theoremstyle{definition}

\newcommand{\off}[1]{}

\usetikzlibrary{matrix,decorations.pathreplacing}
\usetikzlibrary{arrows.meta}

\definecolor{DarkGreen}{rgb}{0.1,0.5,0.1}
\definecolor{DarkRed}{rgb}{0.5,0.1,0.1}
\definecolor{DarkBlue}{rgb}{0.1,0.1,0.5}
\definecolor{DarkPurple}{rgb}{0.5,0.2,0.5}
\definecolor{DarkTurquoise}{rgb}{0.1,0.5,0.5}

\definecolor{beaublue}{rgb}{0.74, 0.83, 0.9}
\definecolor{coolblack}{rgb}{0.0, 0.18, 0.39}
\definecolor{apricot}{rgb}{0.98, 0.81, 0.69}
\definecolor{burntorange}{rgb}{0.8, 0.33, 0.0}
\definecolor{blue-violet}{rgb}{0.54, 0.17, 0.89}
\definecolor{byzantium}{rgb}{0.44, 0.16, 0.39}
\definecolor{brilliantrose}{rgb}{1.0, 0.33, 0.64}
\definecolor{cerisepink}{rgb}{0.93, 0.23, 0.51}
\definecolor{cobalt}{rgb}{0.0, 0.28, 0.67}
\definecolor{bostonuniversityred}{rgb}{0.8, 0.0, 0.0}

\usepackage{mathtools}

\acrodef{dnn}[DNN]{deep neural network}
\acrodef{ml}[ML]{machine learning}
\acrodef{rtt}[RTT]{round-trip time}
\acrodef{rnn}[RNN]{recursive neural network}
\acrodef{lstm}[LSTM]{long short-term memory}
\acrodef{ge}[GE]{Gilbert-Elliott}
\acrodef{snr}[SNR]{signal to noise ratio}
\acrodef{mse}[MSE]{mean squared error}

\usepackage[english]{babel}
\usepackage[utf8x]{inputenc}
\usepackage{amsmath}
\usepackage{graphicx}
\usepackage[colorinlistoftodos]{todonotes}
\usepackage{algorithm}
\usepackage{algpseudocode}
\usepackage{tikz}
\usetikzlibrary{tikzmark,calc}

\newcommand\DrawBox[3][]{%
  \begin{tikzpicture}[remember picture,overlay]
    \fill[gray!10,#1]
    ([xshift= -7.2 em,yshift= 1.9ex]{pic cs:#2})
    rectangle
    ([xshift= 1 em,yshift=-0.5ex]pic cs:#3);
  \end{tikzpicture}%
}

\newcommand\DrawBoxa[3][]{%
  \begin{tikzpicture}[remember picture,overlay]
    \fill[gray!10,#1]
    ([xshift= -6 em,yshift= 1.9ex]{pic cs:#2})
    rectangle
    ([xshift= 2 em,yshift=-0.5ex]pic cs:#3);
  \end{tikzpicture}%
}

\newcommand\DrawBoxb[3][]{%
  \begin{tikzpicture}[remember picture,overlay]
    \fill[gray!10,#1]
    ([xshift= -1.8 em,yshift= 1.9ex]{pic cs:#2})
    rectangle
    ([xshift= 0.3 em,yshift=-0.5ex]pic cs:#3);
  \end{tikzpicture}%
}

\newcommand\DrawBoxc[3][]{%
  \begin{tikzpicture}[remember picture,overlay]
    \fill[gray!10,#1]
    ([xshift= -1.6 em,yshift= 1.9ex]{pic cs:#2})
    rectangle
    ([xshift= 7.5 em,yshift=-0.5ex]pic cs:#3);
  \end{tikzpicture}%
}

\newcommand\DrawBoxd[3][]{%
  \begin{tikzpicture}[remember picture,overlay]
    \fill[gray!10,#1]
    ([xshift= -1.6 em,yshift= 1.9ex]{pic cs:#2})
    rectangle
    ([xshift= 3.8 em,yshift=-0.5ex]pic cs:#3);
  \end{tikzpicture}%
}
\ifCLASSINFOpdf
\else
\fi
\begin{document}
%
\title{\vspace{-0.4cm}Broadcast Approach Meets Network Coding\\
for Data Streaming
\vspace{-0.4cm}}
%
%
%

\author{%
   \IEEEauthorblockN{Alejandro Cohen \IEEEauthorrefmark{1},
                     Muriel M\'edard \IEEEauthorrefmark{2},
                     and Shlomo Shamai (Shitz)\IEEEauthorrefmark{1}}
   \IEEEauthorblockA{\IEEEauthorrefmark{1}%
                      Faculty of Electrical and Computer Engineering, Technion, Israel, \{alecohen, sshlomo\}@technion.ac.il}
    \IEEEauthorblockA{\IEEEauthorrefmark{2}%
                      RLE, MIT, Cambridge, MA, USA, medard@mit.edu}
\vspace{-1.0cm}}

\maketitle


\maketitle

\begin{abstract}
For data streaming applications, existing solutions are not yet able to close the gap between high data rates and low delay. This work considers the problem of data streaming under mixed delay constraints over a single communication channel with delayed feedback. We propose a novel layered adaptive causal random linear network coding (LAC-RLNC) approach with forward error correction. LAC-RLNC is a \emph{variable-to-variable coding} scheme, i.e., variable recovered information data at the receiver over variable short block length and rate is proposed. Specifically, for data streaming with base and enhancement layers of content, we characterize a high dimensional throughput-delay trade-off managed by the adaptive causal layering coding scheme. The base layer is designed to satisfy the strict delay constraints, as it contains the data needed to allow the streaming service. Then, the sender can manage the throughput-delay trade-off of the second layer by adjusting the retransmission rate a priori and posterior as the enhancement layer, that contains the remaining data to augment the streaming service's quality, is with the relax delay constraints. We numerically show that the layered network coding approach can dramatically increase performance. We demonstrate that LAC-RLNC compared with the non-layered approach gains a factor of three in mean and maximum delay for the base layer, close to the lower bound, and factor two for the enhancement layer.

\end{abstract}
\begin{IEEEkeywords}
Broadcast approach, ultra-reliable low-latency, network coding, in-order delivery delay, throughput.
\end{IEEEkeywords}

\vspace{-0.3cm}
  \section{Introduction}\label{sec:intro}
One major challenge in modern communication systems is closing the gap between high data rates and low delay for\off{ real-time} data streaming applications \cite{8329618,anand2018resource}.\off{ This gap is caused due to the high channel variations in the advanced communication networks.} Although high data rates can be achieved by information-theoretic solutions using coding over asymptotic block regime, low delay in ultra-reliable low-latency communications (URLLC) requires short blocks. Therefore, classical information-theoretic coding schemes do not provide the desired throughput-delay trade-offs.

Different approaches have been considered\off{ in recent years} to increase data rates and close this gap. One approach, known as the \emph{Broadcast Approach} \cite{shamai1997broadcast,shamai2003broadcast,tajer2021broadcast}, considers the situation of variable recovered layered information data over a fixed block length, coined \emph{variable-to-fix coding}\off{ as characterized in} \cite{verdu2010variable}. In the classical broadcast channel\off{ fist considered in}, a single sender transmits layered encoded data to a number of receivers \cite{cover1972broadcast,cover2012elements}. Each receiver, obtaining a possibly different channel condition, decodes as many encoded layers as the current channel realization allow it. The broadcast approach realizes the same fundamental principles.\off{However, now} The sender using multi-layering virtually broadcasts all the layers to a single receiver, which can decode layers according to the actual channel condition. For erasure broadcast channels, which can be particularly considered as a degraded broadcast channel in the setting considered herein, time-sharing over long transmissions may be capacity-achieving \cite{korner1977general,boucheron2000priority}. However, in a setting involving data rates and reliability, time-sharing for general channels and even for erasure channels is not optimal \cite{shamai2003broadcast,nair2009capacity}. Moreover, although the broadcast approach can increase data rate and is also considered under mixed delay constraints \cite{whiting2006broadcasting,steiner2008multi,cohen2012broadcast,nikbakht2019mixed,nikbakht2020multiplexing}, solutions using this approach, assuming long transmission blocks with a fixed size that cannot satisfy the low delays requirements for modern URLLC applications.

Another approach of rateless and streaming codes proposed to close this gap \cite{joshi2014effect,cloud2015coded,joshi2016efficient,yang2014deadline,TomFitLucPedSee2014,li2016random,wunderlich2017caterpillar,7117455,fong2019optimal,9076631,9245536,gabriel2018multipath,d2021post,emara2021low,9433517}. Those solutions can be considered as \emph{fix-to-variable coding} schemes, as they consider the scenario of fixed recovered information data over variable block length and rate.\off{ In the presence of delayed feedback, several coding solutions are proposed recently to reduce delay for data streaming \cite{joshi2014effect,cloud2015coded,joshi2016efficient}.} Recently, proposed an \emph{adaptive and causal random linear coding} (AC-RLNC) approach, applied to single-path \cite{9076631}, multi-path and multi-hop \cite{9245536}, and heterogeneous networks with multi-sources and destinations managed by software-defined networking (SDN) \cite{9433517}. AC-RLNC aims at mitigating the throughput-delay trade-off of non-layered data by adapting the required coded retransmissions using a sliding window with a priori and a posteriori forward error correction (FEC) mechanisms. This adaptive and causal adaptation is made by the sender, who tracks the actual channel condition as represented by the delayed feedback acknowledgments. Although AC-RLNC can reduce throughput-delay trade-off, due to the unknown channel realizations at the sender during the delayed feedback, it cannot close this gap\off{ to reach the desired performance}.

In this work, we propose a novel layered adaptive causal network coding scheme for data streaming communications\off{ with mixed delay constraints} coined LAC-RLNC. The proposed approach combines adaptive causal network coding with the broadcast approach to obtain a new \emph{variable-to-variable coding} scheme. That is, in the proposed coding scheme, we examine a new solution of variable recovered information data over variable short block length and rate under mixed delay constraints. Layering\off{ of} data with different priorities using an adaptive\off{ and} causal coded approach opens a new research area to explore that\off{ may achieve} offers a new regime of a high dimensional trade-off of layers and significant gains in delay and throughput guarantees. Specifically, we propose a layering coding scheme for base and enhancement data streaming. The base layer (Layer 1) contains the amount of data needed to allow the streaming service at the receiver in real-time, and the enhancement layer (Layer 2) contains the remaining data that can augment the streaming service's quality. In the adaptive causal solution, the high dimensional trade-off is first managed by adjusting the layering code using the tracking of the current channel condition. The base layer is designed to satisfy the strict delay constraints as it contains the data needed to allow the streaming service. Then, the sender can manage the throughput-delay trade-off of the second layer by adjusting the retransmission rate a priori and posterior as the enhancement layer\off{, containing the remaining data that can augment the streaming service’s quality,} is with relaxed delay constraints. The proposed solution enables layered transmissions with zero error probability under delay constraints \cite{sahai2008block,polyanskiy2011feedback,mary2016finite}.

We contrast the performance of the proposed approach with that of the non-layered AC-RLNC \cite{9076631}. We show that the proposed LAC-RLNC achieves significant gain in delay terms as obtaining about similar data rate, the mean and max in-order delays are reduced approximately by a factor of 3 and a factor of 2, for layer 1 and 2, respectively. Moreover, in the simulated scenarios, we show that the in-order delays of the first layer almost achieve the optimal lover-bound.

The structure of this work is as follows. In Section~\ref{sec:sys}, we formally describe the system model and the metrics in use. In Section~\ref{sec:back}, we provide a background on adaptive causal network coding and on broadcast approach. In Section~\ref{sec:CA}, we present the layered adaptive causal network coding algorithm. In Section~\ref{sec:evaluation}, we evaluate the performance of the proposed solution. Finally, we conclude the paper in Section~\ref{sec:conclusions}.

\begin{figure}
    \centering
    \vspace{-0.4cm}
    \includegraphics[trim=1cm 7cm 1cm 3.8cm,clip,scale=0.4]{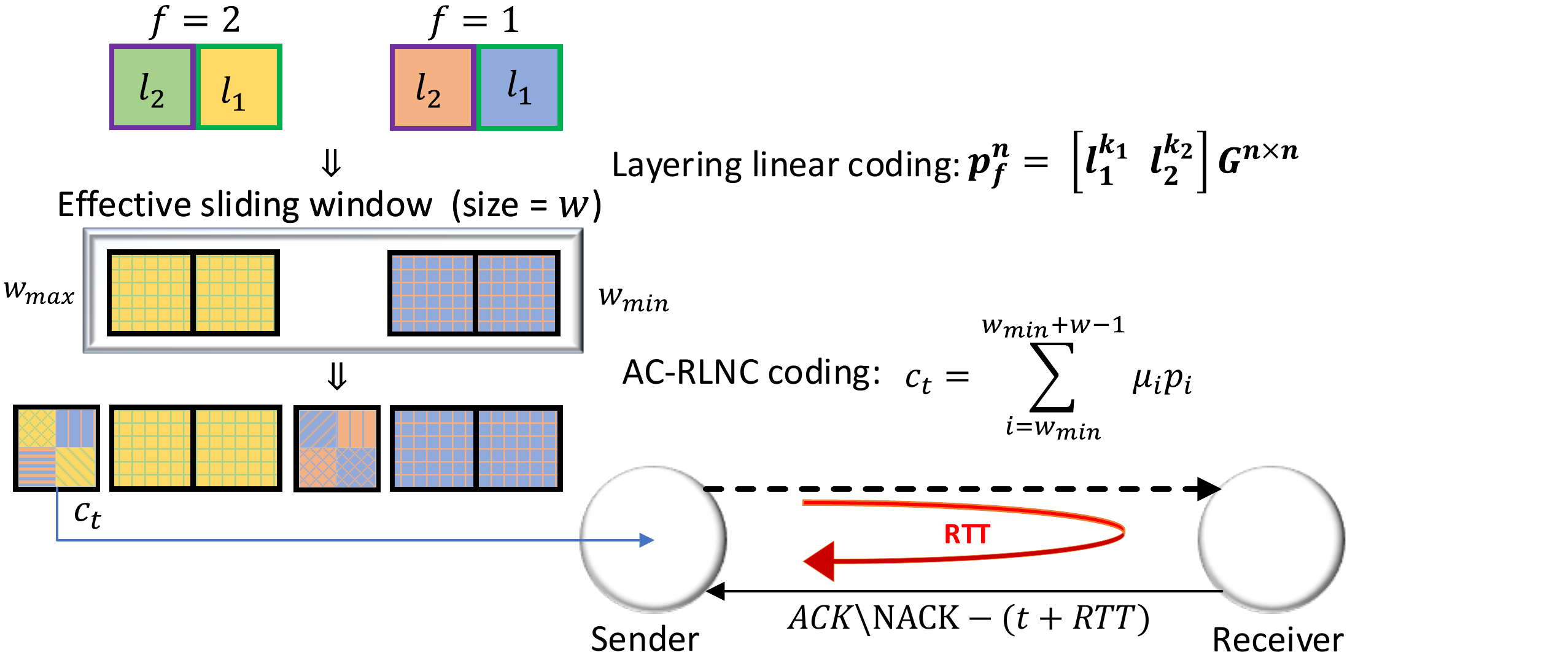}
    \caption{System model and encoding process of LAC-RLNC. The adaptive and causal encoding process with the effective window size $w$ are detailed in Subsection~\ref{subsec:AC-RLNC}. The Layered approach is detailed in Section~\ref{sec:CA}. In this example, for simplicity of notation $w_{\min}=1$.}
	\label{fig:window_coding}
    \vspace{-0.5cm}
\end{figure}

\section{System Model and Problem Formulation}\label{sec:sys}
We consider a point-to-point real-time slotted communication setting with a single forward link and feedback. Fig.~\ref{fig:window_coding} shows the system model. The forward link can be considered as a binary erasure channel (BEC) varying over time with average erasure probability $\epsilon^{\text{mean}}$. The sender is interested in transmitting frames of raw data that can be divided into base layer (Layer 1) and enhancement layer (Layer 2). \off{The base layer (Layer 1) contains the amount of data needed to allow the streaming service at the receiver in real-time, and the enhancement layer (Layer 2) contains the remaining data that can augment the streaming service's quality.} At each time slot $t$, the sender transmits over the forward link a coded packet $c_{t}$ that belongs to the $f$-th frame. The size of the frames is of $n$ coded packets, $k_1$ packets for Layer 1, and the remaining $k_2 \leq n-k_1$ packets for Layer 2. According to the erasure realizations, the receiver sends at each time slot either an acknowledgment (ACK) or a negative-acknowledgment (NACK) message to the sender, using the feedback link. For simplicity, feedback messages are assumed to be reliable, i.e., without errors. The delay between the transmission of a coded packet and the reception of the corresponding feedback is called round trip time (RTT). Hence, for each transmitted coded packet $c_{t}$, the sender receives feedback after $t+ {\rm RTT}$ time slots. Let $u_t$ denote the binary feedback received at time $t$, where
\begin{equation*}
    u_t \triangleq
    \begin{cases}
     1 & \text{received ACK for } c_{t^-},\\
     0 & \text{received NACK at time } c_{t^-},
    \end{cases} \quad t^-\triangleq t-{\rm RTT}.
\end{equation*}
Our main performance metrics are defined as follows:

\begin{table}[t!]
\centering
\begin{tabular}{|l|l|l|}
\hline
{\bf Parameter} & {\bf Definition} \\
    \hline
    \hline
$f$ & frame number\\
$k_i,i \in \{1,2\}$ & number of raw information packets in $i$-th layer\\
$n$ & number of layered coded packets in frame\\
$\textbf{l}_{i}^{k_i}, i \in \{1,2\}$ & raw information packets of $i$-th layer\\
$\textbf{p}_{f}^{n}$ & $n$ layered linear coded information packets\\
    \hline
    \hline
$\eta$ & normalized throughput\\
$D_{\text{mean}}^{i}$ & mean in order delivery delay of $i$-th layer\\
$D_{\text{max}}^{i}$ & maximum in order delivery delay of $i$-th layer\\
    \hline
    \hline
$t$ & time slot index  \\
$c_t$ & coded packet to transmit at time slot $t$ \\
$\mu_i \in \mathbb{F}_z$ &  random coefficients in field $\mathbb{F}_z$\\
$\epsilon_{\text{mean}}$, $\epsilon_{\max}$ & mean and maximum estimated erasure probability \\
$md_t$ & number of DoFs needed to decode $c_t$ \\
$ad_t$ & DoF added to $c_t$  \\
$d$ & rate of DoF ($md_t/ad_t$) \\
$r_{t^-}$ & estimated rate of the channel\\
$th$ & throughput-delay trade-off parameter\\
$\Delta_t\triangleq r_{t^-} - d -th $ & DoF rate gap\\
$\Delta_t >0$ & retransmission criterion \\ 
$\rm RTT$ & round-trip time \\
$w_{\min}$ & index of the first information packet in $c_t$\\
$m_t$ & number of FEC to add per frame \\
EW & end window of $n$ new packets \\
$\overline{o}$ & maximum number of overlap information packets\\
$w\in\{1,\hdots,\overline{o}\}$ & effective window size \\
\hline
\end{tabular}
\vspace{0.0cm}
\caption{LAC-RLNC algorithm: symbol definitions.}
\label{fig:table}
\vspace{-0.6cm}
\end{table}

\noindent (1) {\bf Normalized throughput, $\eta$}. This is defined as the total amount of information data delivered to the receiver in units of bits per second, divided by the total amount of bits transmitted by the sender in this period. 

\noindent (2) {\bf In-order delivery delay of layers}, $D^{i}, i \in \{1,2\}$. This is the difference between the time slot in which an information packet in the $i$-th layer is first transmitted\off{ by the sender} and the time slot in which the layer is decoded in order by the receiver.

Our goal is to maximize the throughput, $\eta$, while minimizing the in-order delivery delay of the layers, $D^1$ and $D^2$.

\section{Background}\label{sec:back}
In this section, we review relevant background in single path adaptive and causal network coding \cite{9076631} and in  broadcast approach \cite{tajer2021broadcast}.

\subsection{Adaptive and Causal Network Coding}\label{subsec:AC-RLNC}

In AC-RLNC, the sender decides at each time step whether to transmit a new coded linear combination or to repeat the last sent combination according to the feedback information. Here, “same” and “new” refer to the raw information packets of information contained in the linear combination. Sending the same linear combination thus means that the raw information packets are the same but with different random coefficients. Let $\mu_i$ and $p_i$ denote the random coefficients drawn from a sufficiently large field and the raw information packets, respectively. Using a sliding window mechanism the coded linear combination transmitted, called a degree of freedom (DoF), given by
\begin{equation}\label{eqn:dof}
    c_{t}=\sum_{i=w_{\min}}^{w_{\max}}\mu_{i}p_{i},
\end{equation}
where $w_{\min}$ corresponds to the oldest raw information packet that is not yet decoded, and $w_{\max}$ is incremented each time a new raw information packet is decided to be included in the linear combination by the sender. We denote by $\text{DoF}(c_t)$ the number of information packets contained in $c_t$.

In this adaptive approach, the sender uses $u_t$ to track the average channel erasure probability $\epsilon_{t}^{\text{mean}}$, and the number of erased and repeated DoFs, denoted $md$ and $ad$, respectively. These tracked quantities are used by two suggested forward error correction (FEC) mechanisms, a priori and a posteriori, to counteract the channel erasures. The a priori mechanism transmits $\lceil\epsilon_{t}^{\text{mean}}\cdot n \rfloor$ repeated DoFs, with $\lceil \cdot \rfloor$ denoting rounding to the nearest integer, periodically after $n$ transmissions of new packets of information. Note that in AC-RLNC given in \cite{9076631}, there are no layering. In the a posteriori mechanism, a retransmission criterion is used by the sender. As demonstrated in \cite{9076631,9245536}, when the estimated channel rate denoted $r_t\triangleq 1-\epsilon_t$ is higher than the rate of the DoFs $d \triangleq md/ad$, the decoder has sufficient DoFs to immediately decode the delivered packets. However, these quantities cannot be computed exactly at the sender due to the \ac{rtt} delay. At time step $t$, the sender can only compute these quantities for time step $t^{-}=t-{\rm RTT}$, using the delayed feedback. Hence, with a tunable parameter $th$, the DoF rate gap and the retransmission decision at each time step, are given respectively by
\begin{equation}\label{eq:cre}
    \Delta_t \triangleq 1 - \epsilon_{t^-}^{\text{mean}} - d -th, \quad \text{and} \quad \Delta_t>0.
\end{equation}
\off{where retransmission is suggested at each time step for which
\begin{equation}\label{eq:re_deci}
    \Delta_t>0.
\end{equation}}
The estimation of the erasure probability $\epsilon_{t^{-}}$  is given by
\begin{equation}\label{e_est}
    \epsilon_{t^{-}}^{\text{mean}} = 1 - \frac{\sum_{j=1}^{t^{-}} u_j}{t^{-}}.
\end{equation}
To manage the maximum delay, the effective sliding window size $w$ is defined, such that $w_{\max}-w_{\min} = w$ and is limited to $\overline{o}$. In AC-RLNC as $th$, $\overline{o}$ can be selected to obtain the desired throughput-delay trade-off. When the limit of the window is reached, the sender transmits the same packet until all the information packets in the linear combination transmitted are decoded.  We refer the readers to \cite{9076631,9245536} for further details on the operation of AC-RLNC.

\begin{algorithm}[t!]\small
\begin{algorithmic}[1]
\Statex {\bf Init:}
\Statex AC-RLNC parameters - $th$, $\overline{o}$, $w_{\min}=1$ and $w_{max}=1$
\DrawBoxb{e}{f}
\Statex \tikzmark{e}Layering code parameters - $n$, $k_1$ and $k_2$ as given in \eqref{eq:n} and \eqref{eq:k}\tikzmark{f}
\Statex\hrulefill
\vspace{0.1cm}
\State {\bf Input:} Feedback $u_t$ for $c_{t^-}$
\DrawBoxc{h}{g}
\State\tikzmark{h}Update $\epsilon_{t^-}^{\text{mean}}$ and $\epsilon^{\max}_{t^-}$ as given in \eqref{e_est} and \eqref{e_est_max}\tikzmark{g}
\State Update $md_t$, $ad_t$, and $\Delta_t$ as given in \eqref{eq:cre}
\DrawBoxd{i}{j}
\State\tikzmark{i}Eliminate the layer's packets decoded from the RLNC \tikzmark{j}
\State Update $w_{\min}$
\If{Feedback NACK, $u_t=0$}
        \If{$\Delta_t \leq 0$}
            \If{not EW}
            \DrawBox{a}{b}
            \If{\tikzmark{a} Entire layered frame $f$ is transmitted}
                \State Encode new frame using layering code in \eqref{eq:layering}
            \EndIf
            \State In order transmit $p_i$ packet from the current frame \tikzmark{b}
            \State Update $w_{\max}$
            \Else
            \State Update $m_t$ as given in \eqref{eq:fec_size}
            \State  Transmit RLNC  $m_t$ times using \eqref{eqn:dof}: $ad_t=ad_t+m_t$
            \EndIf
         \Else  
         \State Transmit RLNC using \eqref{eqn:dof}  $ad_t=ad_t+1$
           \If{EW}
           \State Update $m_t$ as given in
           \State Transmit RLNC $m_t$ times using \eqref{eqn:dof}: $ad_t=ad_t+m_t$
           \EndIf
        \EndIf
\ElsIf{Feedback ACK, $u_t=1$}
      \If{EW}
         \State Update $m_t$ as given in \eqref{eq:fec_size}
         \State Transmit RLNC $m_t$ times using \eqref{eqn:dof}: $ad_t=ad_t+m_t$
      \EndIf
      \If{$\Delta_t > 0$}
         \State Transmit RLNC using \eqref{eqn:dof}: $ad_t=ad_t+1$
      \Else
        \DrawBoxa{c}{d}
        \If{\tikzmark{c}Entire layered frame $f$ is transmitted}
            \State Encode new frame using layering code in \eqref{eq:layering}
        \EndIf
         \State In order transmit $p_i$ packet from the current frame \tikzmark{d}
         \State Update $w_{\max}$
      \EndIf
\EndIf
\If{$ w > \overline{o}$}
   \State Transmit RLNC using \eqref{eqn:dof} until DoF($c_t$)$=0$
\EndIf
\Statex \hspace{-0.55cm}\hrulefill
\vspace{0.1cm}
\Statex \hspace{-0.55cm}$^{[1]}$ The shaded parts are for the layering coding, while the remaining
\Statex\hspace{-0.55cm}parts are for the adaptive and causal coding scheme.
 \caption{Layered adaptive and causal RLNC$^{[1]}$.\label{algo:LAC-RLNC}}
\end{algorithmic}
\end{algorithm}

\begin{figure*}[!h]
    \centering
    \vspace{-0.4cm}
    \includegraphics[width = 1 \textwidth]{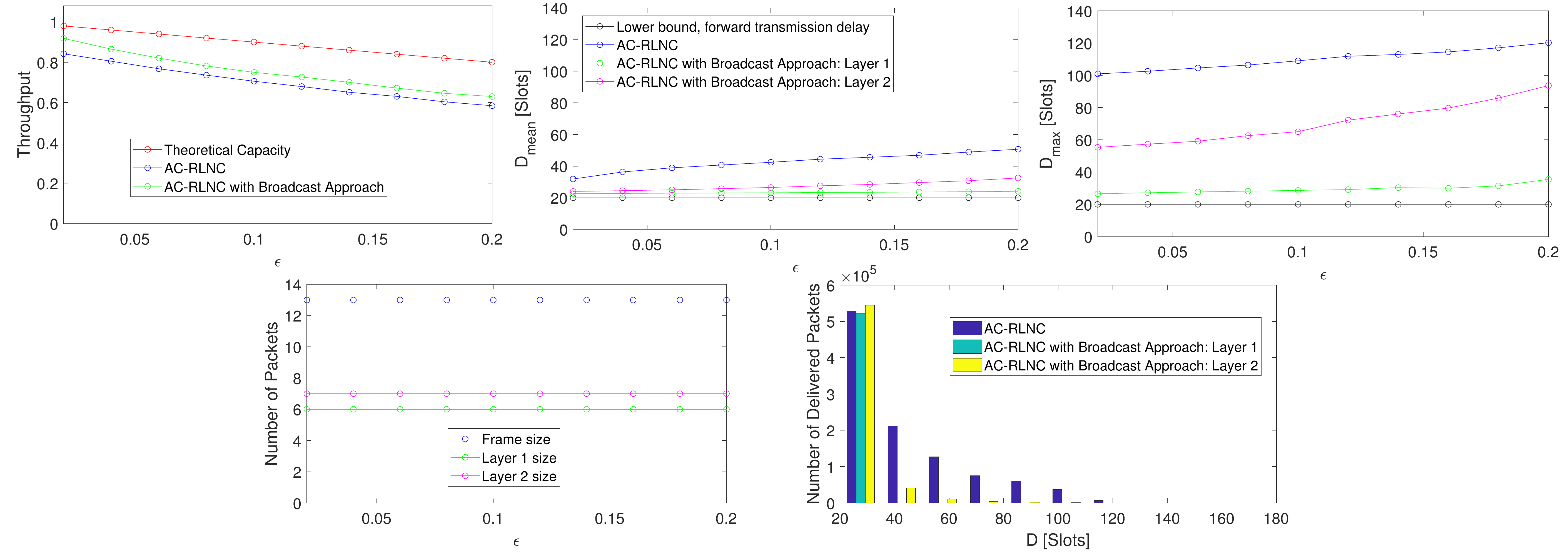}
    \caption{AC-RLNC simulation with and without broadcast approach for \rm RTT = 40 [slots]. The top results are for normalized throughput (left), mean in order delay (middle), and maximum in order delay (right). The bottom results are for the frame, for layer 1 and layer 2 sizes (left), and for the histogram of the in-order delay realizations (right).}
    \label{fig:rtt40}
    \vspace{-0.5cm}
\end{figure*}

\vspace{-0.3cm}
\subsection{Broadcast Approach}\label{subsec:BA}
Layering of raw data was considered in the literature by using Broadcast Approach \cite{shamai1997broadcast,shamai2003broadcast}, to mitigate data rate degradation caused due to the channel temporal variations.\off{To reduce data rate degradation due to the channel temporal variations, layering of raw data was considered in the literature using Broadcast Approach \cite{shamai1997broadcast,shamai2003broadcast}.} In a virtual broadcast channel, the sender encodes multiple ordered layers. Each encoded layer is designed to match the channel observation of a virtual receiver. Then, the sender virtually broadcasts all the layers to the receiver. Thus, the receiver can decode as many layers of layered encoded raw data as the current channel realization allows it. The broadcast approach was considered and characterized for many channel settings and practical solutions. In \cite{tajer2021broadcast} presented an extensive survey of the literature on broadcast approach. Next, we present two coding schemes\off{, from many,} that are considered in the literature for layering\off{ in broadcast approach} setting.

\subsubsection{Random Codes}\label{RLC}
In the literature, random codes, and in particular random linear codes, are known to be capacity-achieving \cite{gallager1973random,domb2015random}. In \cite{cover2012elements}, layering for a broadcast channel with two layers was illustrated by using random code. Due to the space limitation, we refer the readers to \cite[Chapter 15]{cover2012elements} for further details. To significantly reduce the decoding complexity, efficient guessing noise decoders, that can also operate in the short block length regime, has been recently proposed \cite{duffy2018guessing,duffy2019capacity,duffy2019SRGRAND}. Moreover, as elaborated in \cite{duffy2019capacity}, the results in \cite{gallager1973random} using random linear codes may also hold for network coding in heterogeneous networks \cite{effros2003linear,ho2006random}. This motivates future studies on LAC-RLNC for\off{ general} heterogeneous and mesh networks.

\subsubsection{Linear Nested Codes}\label{LNC}
Several nested codes are considered in the literature \cite{subramanian2009mds,zamir1998nested,zamir2002nested,pawar2011securing,zhang2019relaxing}, e.g.. MDS codes \cite{subramanian2009mds}, binary parity-check codes \cite{zamir1998nested}, and lattice codes \cite{zamir2002nested}.\off{In the literature were considered several nested codes \cite{subramanian2009mds,zamir1998nested,zamir2002nested,pawar2011securing,zhang2019relaxing}. For example, using MDS codes \cite{subramanian2009mds}, binary parity-check codes \cite{zamir1998nested}, and lattice codes \cite{zamir2002nested}.}

Let $C_1$ be an $(n,k_1)$ linear code, and $C_2$ be an $(n,k_2)$ linear code with $0 \leq k_1\leq n$ and $0 \leq k_2\leq n-k_1$. Let $\textbf{G}_1$ and $\textbf{G}_2$ be the generator matrices of the codes $C_1$ and $C_2$, respectively. Let $\textbf{l}_i, i\in\{1,2\}$ be the $k_i$ raw data packets of layers 1 and 2, respectively. At the sender, the encoding operation for $n$ packets is given by
\begin{equation}\label{eq:layering}
    \textbf{p}^n = [\textbf{l}_{1}^{k_1}  \textbf{l}_{2}^{k_2}] \left[\begin{array}{c} \textbf{G}_1 \\ \textbf{G}_2 \end{array}\right]
\end{equation}
At the receiver, decoding of the first layer can be performed with $k_1$ delivered encoded packets. The decoding of the second layer can be performed with $n$ delivered encoded packets. Efficient decoding can be done at the receiver, for example, using Reed Solomon (RS) codes as proposed in \cite{subramanian2009mds,zhang2019relaxing}.

\vspace{-0.2cm}
  \section{Layered Adaptive and Causal Coding}\label{sec:CA}
In this section, we detail the layered adaptive and causal RLNC (LAC-RLNC) approach as summarised in  Algorithm~\ref{algo:LAC-RLNC}. We propose a novel solution that benefits from both, Broadcast Approach using linear layering coding and AC-RLNC solution, as described in Subsections~\ref{subsec:BA} and~\ref{subsec:AC-RLNC}, respectively. \off{We propose to marge Broadcast Approach using linear layering coding as described in Subsection~\ref{subsec:BA}, with AC-RLNC solution as described in Subsection~\ref{subsec:AC-RLNC}.} Fig.~\ref{fig:window_coding} shows the system model and the layered adaptive causal encoding process\off{ of the LAC-RLNC protocol with two layers and a sliding window mechanism}. The symbol definitions\off{ of LAC-RLNC} are provided in Table \ref{fig:table}.

The proposed LAC-RLNC approach is designed to minimize the in-order delivery delay of the base layer (Layer 1) while closing the throughput-delay trade-off of the enhancement layer (Layer 2). The sander decides at each time step whether to transmit an FEC RLNC or new information coded packet using a sliding window mechanism as in AC-RLNC. However, the proposed layered approach differs from AC-RLNC in two main perspectives: (1) Here, the raw data is first encoded using a broadcast approach coding that allows the receiver to decode the base layer (Layer 1) with minimum in-order delay required. At the same time, the enhancement layer (Layer 2) with the a priori and posterior FEC mechanisms allow managing the desired throughput-delay trade-off. (2) RLNC coding is used to transmit a repeat layered coded packet using a linear combination with random coefficients as given in \eqref{eqn:dof}. That is, RLNC is used only for FEC retransmissions as controlled by the a priori and posterior mechanisms, not for all the coded packets transmitted as described\off{ for AC-RLNC} in Section~\ref{subsec:AC-RLNC}. The components of the proposed protocol are described next.

\begin{figure*}[!h]
    \centering
    \vspace{-0.4cm}
    \includegraphics[width = 1 \textwidth]{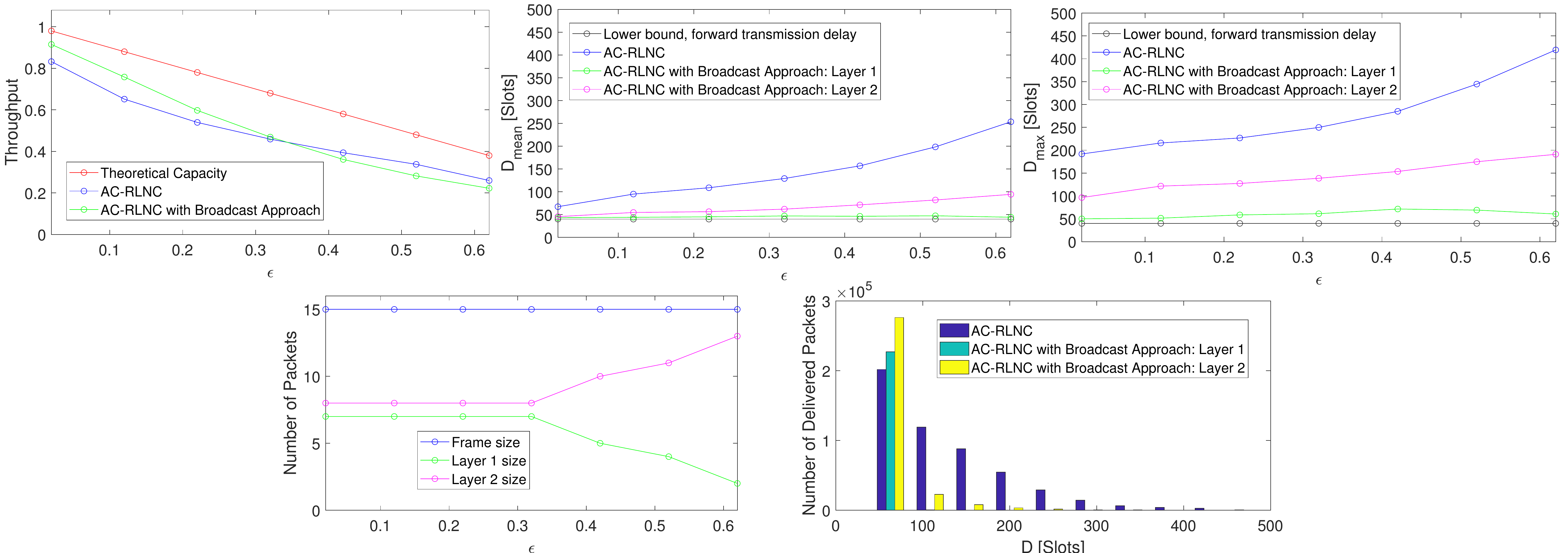}
    \caption{AC-RLNC simulation with and without broadcast approach for \rm RTT = 80 [slots]. The top results are for normalized throughput (left), mean in order delay (middle), and maximum in order delay (right). The bottom results are for the frame, for layer 1 and layer 2 sizes (left), and for the histogram of the in-order delay realizations (right).}
    \label{fig:rtt80}
    \vspace{-0.5cm}
\end{figure*}

\paragraph{Layered Coded Packets} The raw data packets are first encoded in-order into frames of base and enhancement layers  (resp. to lines 9-11 and 31-34 in Algo.~\ref{algo:LAC-RLNC}) using the layering code as described in Section~\ref{LNC}. The code and layers sizes are selected to obtain the desired high dimensional throughput-delay trade-off. The frame size is given by
\begin{equation}\label{eq:n}
    n = {\rm RTT}/n_f,
\end{equation}
where $n_f$ denotes the frame size factor, which is a tunable parameter. The code designer can control the minimum in-order delivery delay at the receiver by varying this factor. The layers' packets number are given by
\begin{equation}\label{eq:k}
    k_1 = n(1-\epsilon_{\max}) \quad \text{  and  } \quad k_2 = n-k_1
\end{equation}
where $\epsilon_{\max}$ denotes the maximum erasure probability estimated, as given in \eqref{eq:era}. The packets number are selected by the code designer to manage the layer 1 throughput-delay trade-off. The maximum erasure probability estimated is given by
\begin{equation}\label{eq:era}
    \epsilon_{\max} = \epsilon_{\text{mean}} + \sigma\frac{\sqrt{v}}{n}, \quad \text{where} \quad
    v = n(1-\epsilon_{\text{mean}})\epsilon_{\text{mean}}
\end{equation}
is the variance of the BEC during the period of n transmissions, and $\sigma$ is the tunable standard deviation factor parameter, using the so-called $68 - 95 - 99.7$ rule. Note that the layer's size can be adjusted adaptively according to the tracking of the channel, as represented by the feedback. 

\paragraph{Effective Window}
At each time step, the sender tracks the actual channel and DoF rates by the feedback acknowledgments $u_t$, and updates the DoF rate gap (resp. to lines 1,2,3 in Algoo.~\ref{algo:LAC-RLNC})\off{At each time step, by the feedback acknowledgments $u_t$, the sender tracka the actual channel and the DoF rates, and updates the DoF rate gap (resp. to lines 1, 2, and 3 in Algo.~\ref{algo:LAC-RLNC})}. In LAC-RLNC, decoding at the receiver is done for coded layers, not per coded packets. The effective window, including the subset of information packets within the effective window, is updated for the decoded layer at the receiver (resp. to lines 4 and 5 in Algo.~\ref{algo:LAC-RLNC}), as reflected at the sender by the delayed feedback. That is, $w_{\text{min}}$ and the information packets needed in the coded RLNC are updated when $k_1$ or $k_1+k_2$ coded packets for each frame are delivered at the receiver and provide sufficient DoF's to be decoded.

\paragraph{Layered FEC mechanisms}
We use both FEC mechanisms, a priori or a posteriori, as given in Subsection~\ref{subsec:AC-RLNC}. The sender encodes the layered coded packets to RLNC packet for any FEC retransmission, as given in \eqref{eqn:dof}, using the effective window. We use the retransmission criteria as given in \eqref{eq:cre} (resp. to lines 19 and 31 in Algo. 1) for the a posteriori mechanism decisions.\off{For any FEC retransmission, the sender encoded the layered coded packets to RLNC packet as given in \eqref{eqn:dof} using the effective window. For the a posteriori mechanism we use the retransmission criteria as given in \eqref{eq:cre}\off{ and \eqref{eq:re_deci}} (resp. to lines 19 and 31 in Algo.~\ref{algo:LAC-RLNC}).} Finally, to obtain the desired high dimensional throughput-delay trade-off, the adaptation of the layered a priori mechanism is designed as follows: The adaptive number of the periodically a priori RLNC retransmission $m_t$, after transmission of each frame, is given by
\begin{equation}\label{eq:fec_size}
    m_t = \lceil \epsilon^{\max}_{t^-} \cdot n\rfloor
\end{equation}
where
\begin{equation}\label{e_est_max}
    \epsilon^{\max}_{t^-} = \epsilon_{t^-}^{\text{mean}} + \sigma_m\frac{\sqrt{v}}{n},
\end{equation}
and $\sigma_m$ is selected to control the trade-off of layer 2.

  \section{Performance Evaluation}\label{sec:evaluation}

This section presents the results of the layered AC-RLNC approach for single path communication with feedback. In terms of mean and max in-order delivery delay and throughput, the performance is compared to the non-layered solution.

The simulation results presented are for a BEC channel. The first scenario we consider is for erasure probability varying between 0.02 to 0.2 and with $\rm RTT$ of 40 [Slots]. The second scenario is for erasure probability varying between 0.02 to 0.6 and with $\rm RTT$ of 80 [Slots].  Figures~\ref{fig:rtt40} and~\ref{fig:rtt80}, show the results of the first and second scenarios, respectively, using AC-RLNC, with and without layering. The top results are for the normalized throughput $\eta$ (left), and mean $D_{\text{mean}}$ (middle) and max $D_{\text{max}}$ (right) in-order delivery of frames, as defined in Section~\ref{sec:sys}. The results presented on the top, have been averaged on 150 different channel realizations. The bottom results are of the frame and the layer's sizes selected to obtain the desired high dimensional throughput-delay trade-off (left) and the histogram of the in-order delay realizations (right).

In both scenarios, the layering code parameters are selected to obtain similar performance as the non-layered AC-RLNC solution in terms of throughput with zero error probability. Thus, we compare the performance, of the layered and non-layered solutions, in terms of in-order delivery delay of frames. We can see significant gains in $D_{\text{mean}}$ and $D_{\text{max}}$ with respect to the non-layered solution. From the delay point of view, the gains are significant as both the mean and max in-order delay are reduced approximately by a factor of 3 and a factor of 2 for layers 1 and 2, respectively. Moreover, for each scenario, given the $\rm RTT$ delay, we show the lower bound of the delay. For relatively good forward link quality, this lower bound reducing to about half of the $\rm RTT$ \cite{9245536}. We note that with the selected layering code parameters, the in-order delivery delay of layer 1 is very close to the optimal lower bound. In the histogram figures of the in-order devilry delay, we can see that all the realizations of layer 1 are obtained close to the optimal lower bound delay. This is critical for streaming applications, as by using the proposed approach, the base layer can be delivered in a minimum delay to enable the required service. Moreover, we can note that using the Broadcast approach, more than $90\%$ realizations of the enhancement layer are almost achieving the\off{ optimal} lower bound.

  \section{Conclusions and Future Work}\label{sec:conclusions}
We proposed a layered adaptive causal RLNC (LAC-RLNC) algorithm for data streaming over an erasure channel. The proposed method considered variable recovered layered information over variable short block length and rate. Specifically, we show that with zero error probability LAC-RLNC can reduce the in-order delay of the layers significantly and design the high dimensional gap between data rates and delay to obtain the desired performers.

Future work includes the derivation of bounds on the mean and maximum in order delivery delay and throughput for the proposed approach. These bounds will provide theoretical guarantees for the layered code designer in the adaptive and causal network coding solution. Extensions also include the study of layered adaptive causal network codes where we have more than two layers and coded layered solutions for multi-path and multi-hop heterogeneous networks.

\off{\DrawBox{a}{b}
\begin{algorithm}
\caption{Expand $K$- Plex}{ }
\begin{algorithmic}[1]
\Statex
\INPUT\tikzmark{a}
\Statex $a$ \Comment The a of the algo
\Statex $b$  \Comment The b of the algo
\Statex $c$  \Comment The c of the algo\tikzmark{b}
\Statex
\State $V \gets 0$
\end{algorithmic}
\end{algorithm}

\DrawBox[draw=orange,fill=orange!30]{c}{d}

\begin{algorithm}
\caption{Expand $K$- Plex}{ }
\begin{algorithmic}[1]
\Statex
\INPUT\tikzmark{c}
\Statex $a$ \Comment The a of the algo
\Statex $b$  \Comment The b of the algo
\Statex $c$  \Comment The c of the algo\tikzmark{d}
\Statex
\State $V \gets 0$
\end{algorithmic}
\end{algorithm}}

\bibliographystyle{IEEEtran}
\bibliography{IEEEabrv,references,Ref1,Ref2}

\end{document}